\documentstyle[a4,amsmath,amssymb,latexsym,epsfig,citesort]{article}
\def \d{{\mathrm{d}}}

\def \pd{\partial}

\def \tl#1{\overset{\kern 1pt\circ}{#1}}
\def \TL#1{\overset{\kern -3pt \circ}{#1}}
\def \TLL#1{\overset{\kern -7pt \circ}{#1}}

\begin{document}
\title{{\bf On the correspondence between 
a screw dislocation in gradient elasticity and 
a regularized vortex}}
\author{{\bf Markus Lazar\footnote{E-mail: lazar@lmm.jussieu.fr}}\\ \\
Laboratoire de Mod{\'e}lisation en M{\'e}canique,\\
        Universit{\'e} Pierre et Marie Curie,\\
	4 Place Jussieu, Case 162,\\	
	F-75252 Paris Cedex 05, France\\
}
\maketitle
\begin{abstract}
We show the correspondence between a screw dislocation 
in gradient elasticity and a regularized vortex. 
The effective Burgers vector, 
nonsingular distortion and stress fields of
a screw dislocation and the effective circulation, smoothed velocity and 
momentum of a vortex are given and discussed. 
\\

\end{abstract}
\vspace*{2mm}

\section{Introduction}
The close relationship between vortices in fluid mechanics and dislocations
in elasticity theory has been known for many years (see, e.g.,~\cite{Nabarro,Marc}). Both the dislocations
and the vortices have singularities at the dislocation line and vortex line, 
respectively. Nowadays it is very popular to regularize the stress and/or strain
fields of a dislocation by means of nonlocal elasticity~\cite{Eringen83,Eringen02} 
and gradient elasticity~\cite{GA96,GA99,GA99b,Lazar02,Lazar03,LM03}. 
In such a theory an additional parameter called gradient coefficient 
is introduced. 
On the other hand, in fluid mechanics a so-called ``alpha-model'' 
is used to smooth out the velocity of a vortex and an arbitrary parameter
$\alpha$ appears~\cite{Holm,MP,Holm2}. 

The main purpose of the present letter will be to give a one to one 
correspondence between the field quantities of a 
nonsingular screw dislocation in gradient elasticity and a smoothed vortex
in fluid mechanics. Some similarities between gradient elasticity and
the so-called ``alpha-model'' will be discussed.   
For instance, we will see that the parameter $\alpha$ is a kind of a 
gradient coefficient in such a model.

\section{Screw dislocation in gradient elasticity}
Dislocations are fundamental line-like defects in solids.
We consider a single screw dislocation contained within an infinitely extended solid body. 
Its dislocation line and Burgers vector are directed 
along the $z$-direction. 
The Burgers vector is the measure of strength (topological charge) 
of a dislocation and may be expressed in elasticity as follows
\begin{align}
b_i=\oint\tl\beta_{ij}\,\d x_j,
\end{align} 
where $\tl\beta_{ij}$ is the elastic distortion tensor in elasticity.
In cylindrical coordinates, the elastic distortion of a single screw 
dislocation is given by~\cite{deWit}
\begin{align}
\label{dist-cl}
\tl\beta_{z\varphi}=
\frac{b}{2\pi r},
\end{align}
where $r^2=x^2+y^2$.  
Therefore, we obtain $b_z=b$.
In gradient elasticity, one may define a smoothed distortion tensor
$\beta_{ij}$ which has to satisfy the following 
inhomogeneous Helmholtz equation
\begin{align}
\label{dist-HE}
\big(1-\kappa^{-2}\Delta\big)\beta_{ij}=\tl\beta {}_{ij},
\end{align}
where the right hand side is given in terms of $\tl\beta {}_{ij}$.
It is obvious that Eq.~(\ref{dist-HE}) may be rewritten in the form of
a integral relation as a convolution integral
\begin{align}
\label{dist-int}
\beta_{ij}(r)=\int G(r-r')\tl\beta_{ij}(r')\d V(r'),
\end{align}
where the two-dimensional Green function is given by
\begin{align}
G(r)=\frac{\kappa^2}{2\pi}\, K_0(\kappa r).
\end{align}
Here $K_n$ denotes the modified Bessel function of the second kind and of 
order $n$.
In Eq.~(\ref{dist-HE}) a parameter $\kappa$ 
with the dimension of an inverse length is introduced. 
Eventually, one may define a characteristic length scale 
$\varepsilon=\kappa^{-1}$. 

Substituting~(\ref{dist-cl}) into Eq.~(\ref{dist-HE}), we find for 
the distortion in gradient elasticity~\cite{GA96,Lazar02,Lazar03} 
\begin{align}
\label{dist-cyl}
\beta_{z \varphi}=\frac{ b}{2\pi}\,\frac{1}{r}\Big\{1-\kappa r K_1(\kappa r)\Big\}.
\end{align}
This distortion is smoothed within the dislocation
core and decays like $r^{-1}$ for large $r$.
It is zero at $r=0$ and has a maximum of
$\beta^{\rm max}_{z\varphi}\simeq 0.399\kappa b /[2\pi]$ 
at $r\simeq 1.114\kappa^{-1}$. 
The distortion~(\ref{dist-cyl}) satisfies the condition, 
$\pd_j\beta_{ij}=0$.

Using this smoothed distortion, the effective Burgers vector is calculated 
as~\cite{Lazar02,Lazar03}
\begin{align}
\label{Burger-screw}
b_z(r)=\oint\beta_{z\varphi} r\, \d \varphi
      =b\Big\{1-\kappa r K_1(\kappa r)\Big\}.
\end{align}
It differs from the constant value $b$ in the 
region from $r=0$ up to $r\simeq 6/\kappa$. 
In fact, we find $b_z(0) =0$ and $b_z(\infty)=b$.

The dislocation density tensor is defined by
\begin{align}
\label{DD}
\alpha_{ij}=\epsilon_{jkl}\pd_k\beta_{il}
\end{align}
and fulfills the Bianchi identity for the torsion
\begin{align}
\pd_j\alpha_{ij}=0,
\end{align}
which means that dislocations cannot end inside the solid body.
With (\ref{dist-cyl}) and (\ref{DD})
the dislocation density of a single screw dislocation is obtained as~\cite{Lazar02,Lazar03}
\begin{align}
\label{alpha-zz}
\alpha_{zz}=
\frac{b\kappa^2}{2\pi}\, K_0(\kappa r).
\end{align}
In the limit as $\kappa^{-1}\rightarrow 0$, 
Eq.~(\ref{alpha-zz}) converts to the classical dislocation density
$\alpha_{zz}=b\,\delta(x)\delta(y)$.

By means of the smoothed distortion and the Hooke law
\begin{align}
\sigma_{ij}=2\mu \beta_{(ij)}+\lambda \delta_{ij}\beta_{ll},
\end{align}
where $\mu$ and $\lambda$ are the Lam{\'e} constants, 
the smoothed stress tensor of the single screw dislocation has
the following form~\cite{Eringen83,Eringen02,GA99,Lazar02,Lazar03,LM03}
\begin{align}
\sigma_{z\varphi}=
\frac{\mu b}{2\pi}\,\frac{1}{r}\Big\{1-\kappa r K_1(\kappa r)\Big\},
\end{align}
which does not possess a singularity.

\section{Regularized vortex}
Vortices are fundamental line-like excitations of a fluid.
We consider a vortex in an infinitely extended fluid  
for which the conditions of static current and of incompressibility are fulfilled
\begin{align}
\pd_t\omega_i=(\omega_j\pd_j)v_i=0,
\qquad\text{with}\qquad
\pd_iv_i=0. 
\end{align}
Here $v_i$ denotes the fluid velocity and $\omega_i=\epsilon_{ijk}\pd_j v_k$ is the 
vorticity vector.
The vortex line lies in the $z$-direction.
The fluid velocity of such a single vortex is in fluid mechanics given by~\cite{Sommer}
\begin{align}
\label{vel-cl}
v_{\varphi}=\frac{\Gamma}{2\pi r},
\end{align}
which is infinite at $r=0$. 
Here, $\Gamma$ denotes the circulation which is a measure of strength 
of the singular vortex and $\Gamma=\oint v_i\d x_i$. 
Therefore, the corresponding vortex density reads 
\begin{align}
\label{VD-cl}
\omega_z=\Gamma\,\delta(x)\delta(y).
\end{align}

\begin{figure}[t]\unitlength1cm
\centerline{
\begin{picture}(9,6)
\put(0.0,0.2){\epsfxsize=9cm\epsffile{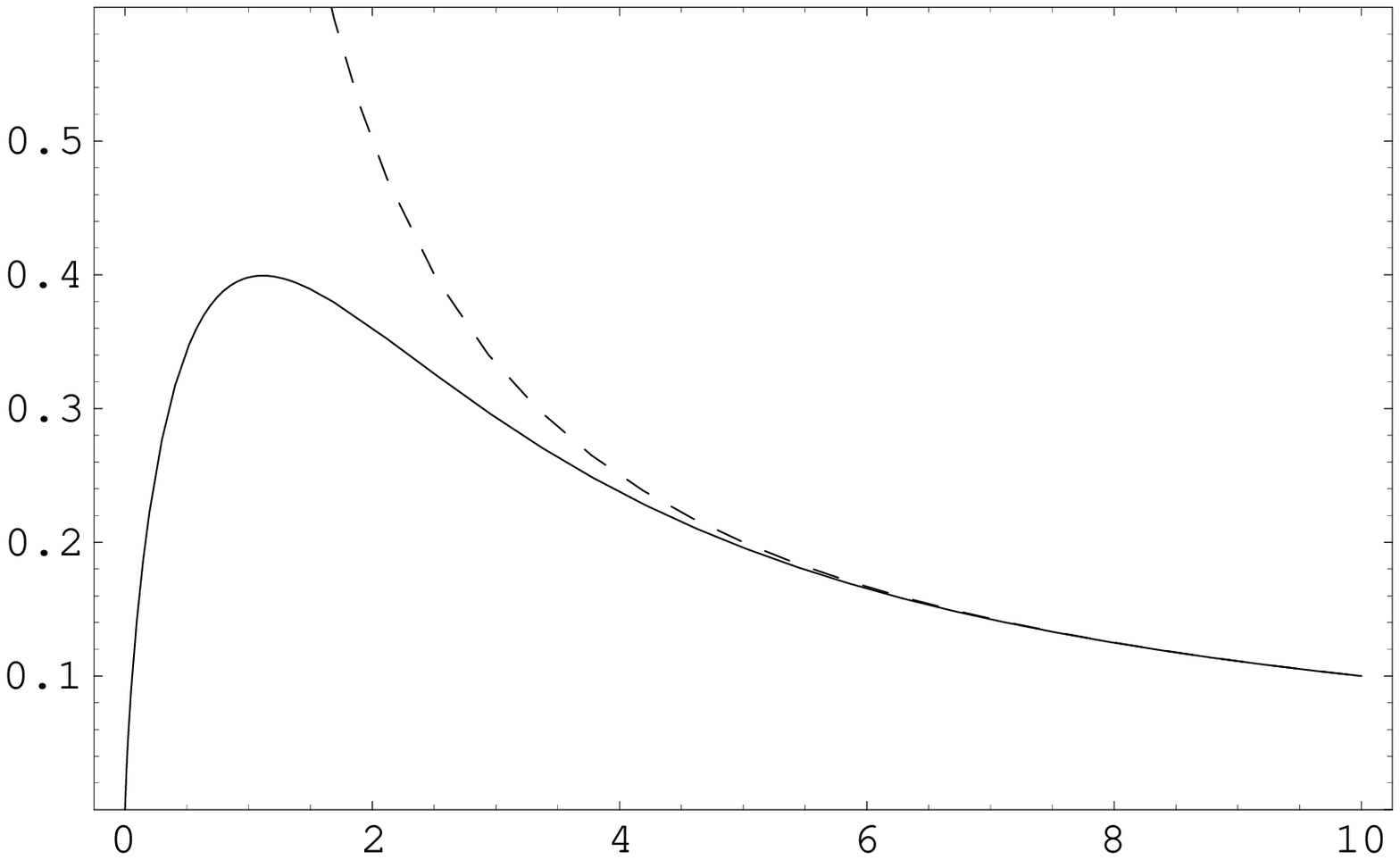}}
\put(4.5,0.0){$r/\alpha$}
\put(-1.0,4.5){$u_{\varphi}$}
\end{picture}
}
\caption{Smoothed velocity $u_{\varphi}$ of a vortex (solid curve)
is given in units of $\Gamma/[2\pi\alpha]$.
The dashed curve represents the classical solution.}
\label{fig:velo}
\end{figure}

On the other hand,
one may define a locally smoothed velocity $u_i$ by
the following inhomogeneous Helmholtz equation (see, e.g.,~\cite{Holm2,MP}) 
\begin{align}
\label{vel-HE}
\big(1-\alpha^{2}\Delta\big) u_{i}=v_{i},
\end{align}
and the corresponding integral relation
\begin{align}
\label{u-int}
u_{i}(r)=\int G_\alpha (r-r') v_{i}(r')\d V(r'),
\end{align}
which is introduced in the ``alpha-model''.
It is interesting to note that Eqs.~(\ref{vel-HE}) and (\ref{u-int}) 
are similar in form as Eqs.~(\ref{dist-HE}) and (\ref{dist-int}), respectively.
For two-dimensional problems $G_\alpha$ is given by
\begin{align}
G_\alpha(r)=\frac{1}{2\pi\alpha^2}\, K_0\big(r/\alpha\big),\qquad
r=\sqrt{x^2+y^2}
\end{align}
and the three-dimensional Green function reads
\begin{align}
G_\alpha(r)=\frac{1}{4\pi\alpha^2 r}\, \exp\big(-r/\alpha\big),\qquad
r=\sqrt{x^2+y^2+z^2}.
\end{align}
In our case the problem has cylindrical symmetry around the vortex line 
and, thus, it is two-dimensional.
Using (\ref{vel-HE}) with (\ref{vel-cl}),
the smoothed velocity for a single vortex is found as
\begin{align}
\label{vel-sm}
u_{\varphi}=\frac{\Gamma}{2\pi}\,\frac{1}{r}\Big\{1-r/\alpha\, K_1\big( r/\alpha\big)\Big\}.
\end{align}
This smoothed velocity is displayed graphically in Fig.~\ref{fig:velo}.
It has no singularity.
In fact,
it is zero at $r=0$ and the maximum value of
$u^{\rm max}_{\varphi}\simeq 0.399 \Gamma /[2\pi\alpha]$ 
occurs at $r\simeq 1.114\alpha$.
Thus, the position and the value of the maximum velocity depend on the 
$\alpha$-parameter. On the other hand, these quantities should be measured in
an experiment or in a related simulation. 
Therefore, in this way one can determine the $\alpha$-parameter for 
such a single vortex.

\begin{figure}[t]\unitlength1cm
\centerline{
\begin{picture}(9,6)
\put(0.0,0.2){
\epsfxsize=9cm\epsffile{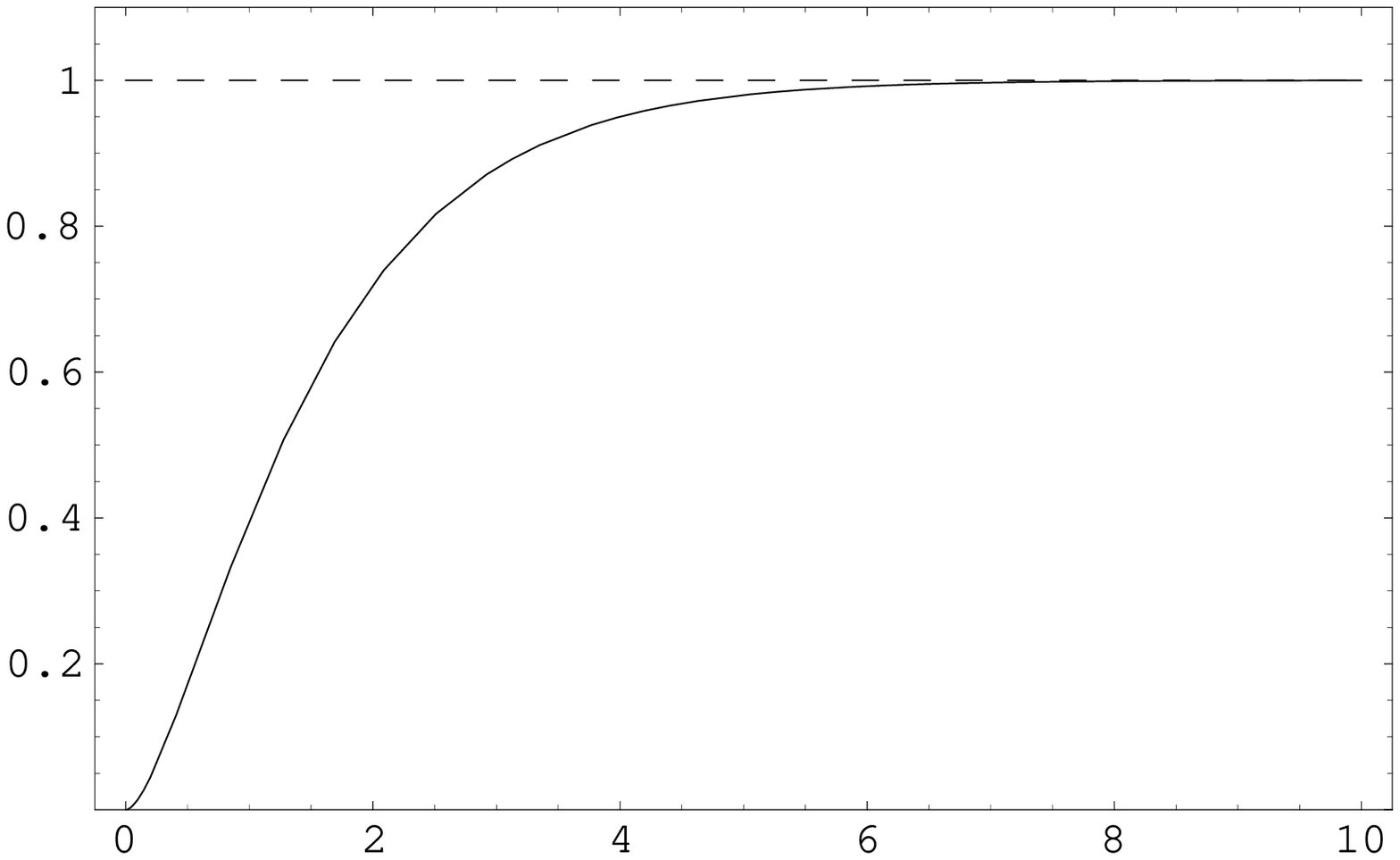}}
\put(4.5,0.0){$r/\alpha$}
\put(-1.7,4.5){$\Gamma(r)/\Gamma$}
\end{picture}
}
\caption{Effective circulation $\Gamma(r)/\Gamma$ (solid curve).}
\label{fig:circ}
\end{figure}

Using Eq.~(\ref{vel-sm}), the effective circulation may be calculated  
\begin{align}
\label{circ-eff}
\Gamma(r)=\oint u_{\varphi} r\, \d \varphi
      =\Gamma\Big\{1-r/\alpha\, K_1\big(r/\alpha\big)\Big\}.
\end{align}
The effective circulation is plotted in Fig.~\ref{fig:circ}.
One can see that
it differs from the constant value $\Gamma$ in the 
region from $r=0$ up to $r\simeq 6\alpha$. 
It has $\Gamma(0) =0$ and $\Gamma(\infty)=\Gamma$.
Thus, one might
take $r_c\simeq 6\alpha$ as the core radius of the vortex.
Again, the $\alpha$-parameter can be determined by the measurement of 
the profile for the effective circulation.

In analogy to the dislocation tensor~(\ref{alpha-zz}),
one may define a smoothed vorticity or vortex density as follows
\begin{align}
\tilde\omega_{i}=\epsilon_{ijk}\pd_j u_{k}
\end{align}
and it fulfills the ``Bianchi identity in fluid mechanics'' 
\begin{align}
\pd_i\tilde\omega_{i}=0.
\end{align}
Thus, vortices cannot end inside the fluid.
With (\ref{vel-sm}) the vortex density is given by
\begin{align}
\label{VD2}
\tilde\omega_z=\frac{\Gamma}{2\pi\alpha^2}\,K_0\big(r/\alpha\big),
\end{align}
which is smeared out around the vortex line.
It is interesting to note that
in the limit as $\alpha\rightarrow 0$, 
$\tilde\omega_z$ approaches $\omega_{z}$ which is given by Eq.~(\ref{VD-cl}).

Eventually, the smoothed momentum vector is defined as 
\begin{align}
\tilde p_i=\varrho u_i,
\end{align}
where $\varrho$ denotes the density. 
For the single vortex we find
the smoothed momentum 
\begin{align}
\tilde p_{\varphi}=\frac{\varrho\Gamma}{2\pi}\,\frac{1}{r}\Big\{1-r/\alpha\, K_1\big( r/\alpha\big)\Big\}.
\end{align}
It does not possess any singularity. 

\section{Summary}
We have shown a one to one relationship between a screw dislocation in 
gradient elasticity and a smoothed vortex.  
A review of the correspondence between a regularized vortex
in fluid mechanics and a screw dislocation in gradient elasticity 
is given in Table~\ref{tab1}.
\begin{table}[h]
\caption{The correspondence between a vortex and a screw 
dislocation}
\begin{center}
\begin{tabular}{llll}
\label{tab1}
\\
\hline\\
vortex  & & screw dislocation &\\ \\
\hline\\
smoothed velocity & $u_\varphi$ &smoothed elastic distortion &$\beta_{z\varphi}$  \\
effective circulation & $\Gamma(r)$ & effective Burgers vector & $b_z(r)$ \\
smoothed circulation & $\tilde\omega_z$ & dislocation density & $\alpha_{zz}$\\
density & $\varrho$ & shear modulus & $\mu$ \\
smoothed momentum & $\tilde p_\varphi$ & smoothed stress field & $\sigma_{z\varphi}$\\
characteristic length scale & $\alpha$ & characteristic length scale & $\varepsilon=1/\kappa$ 
\\ \\
\hline
\end{tabular}
\end{center}
\end{table}
One mathematical distinction, however, is that the characteristic fluid quantities are
vectors or scalars and their counterparts in solids are second rank tensors
or vectors. 
We have discussed the similarities between gradient elasticity and the basic equations
in the ``alpha-model''. It turns out that the ``alpha-model'' may be considered 
as a gradient or nonlocal theory for the vortex in fluid mechanics.
The $\alpha$-parameter is the gradient parameter in the model.
In addition, we have given some hints to determine the $\alpha$-parameter 
for a vortex in an experiment or simulation.

\end{document}